\documentstyle[12pt,epsfig,rotate]{article}

\begin{document}

\begin{titlepage}
 
\pagebreak

\vspace*{0.1cm}
\begin{center}
{\huge Institute of Theoretical }
\end{center} 
\begin{center}
{\huge and Experimental Physics }
\end{center} 
\begin{center}
{\huge Preprint  15-01 }
\end{center} 
\vspace{3.0cm}
\begin{center}
{\large M.Danilov, Yu. Gilitsky, T. Kvaratschellia, L.Laptin, I.Tichomirov,}
\end{center} 
\begin{center}
\vspace{0.05cm}
{\large M.Titov, Yu.Zaitsev}
\end{center} 

\begin{center}
\vspace{2.cm}
\end{center}
\begin{center}
\vspace{0.1cm} 
{\large\bf Aging Studies of Large Area Proportional }
\end{center}
\begin{center}
\vspace{0.1cm} 
{\large\bf Chambers under High-Rate Irradiation }
\end{center}
\begin{center}
\vspace{0.1cm} 
{\large\bf with $CF_4$-based Mixtures (PART 2). }
\end{center}
\begin{center}
\vspace{0.1cm} 
\end{center}
 
\vspace{1.5cm}
\begin{center}
{\large \bf  PART 2 }
\end{center}

\vspace{5.5cm}

\begin{center}
{\large\bf Moscow 2001}
\end{center} 

\end{titlepage}

\newpage

\vspace*{7.5cm}

\large 

 Experimental conditions at the HERA-B experiment impose very strong 
requirements for gaseous detectors.
The charged particle fluxes through the HERA-B tracking system,
varying with the radial distance $R$ from the beam line,
are about $2 \times 10^{7}/R^{2}$ particles per second,
and comparable to those that will be encountered by LHC experiments.

 The severe radiation environment 
of the HERA-B experiment leads to a maximum
charge deposit on a wire, within the muon detector, 
of 200 mC/cm per year. We report recent results of aging studies 
performed by irradiating proportional wire chambers
filled with $Ar/CF_4/CH_4$ (74:20:6), $Ar/CF_4/CH_4$ (67:30:3),
$Ar/CF_4/CO_2$ (65:30:5), $Ar/CF_4$ (70:30), $CF_4/CH_4$ (90:10),
$CF_4/CH_4$ (80:20) mixtures in a three different  experimental setups.
 The size of the irradiation zone varied in the tests 
from 1~cm  up to 500~cm. 
Our experience shows that the aging rate depends not only on the total
collected charge, but, in addition, on the mode of operation and area
of irradiation.
 The possible 
application of these results to the construction of a large area 
gaseous detectors for operation in high rate environments is presented.

\large
\clearpage
\pagenumbering{arabic}
\setcounter{page}{1}

\vspace*{2.2cm}

\section{ Aging in an $Ar/CF_4$(70:30) gas mixture}
    
 In order to study the influence of materials, cathode cleanliness,
and water addition on aging performance 
we have carried out radiation tests with an  $Ar/CF_4$(70:30) gas,
which has very similar transport properties
with $Ar/CF_4/CH_4$ (67:30:3) mixture.
 It is well known that hydrocarbon free mixtures
are much more resistant to polymerization effects
than gases containing hydrocarbons.
 The aging results reported in this paper also
demonstrate that the lifetime of aluminum proportional chambers
changes drastically by replacing 3~$\%$ of $CH_4$ by 5~$\%$
of $CO_2$ in the $Ar/CF_4$-based mixture. 
 Thus, this difference between aging properties can be solely attributed
to the change in the gas phase and gas surface chemical reactions
in the wire avalanches, as a result of
the removal of $CH_4$ and addition $CO_2$ to the mixture. 
From the viewpoint of plasma polymerization,
the main effect of the $CF_4$ plasma 
can be shifted from etching to polymer deposition by the 
addition of $H_2$ into the plasma and that depending 
upon the amount of $H_2$ added to $CF_4$, the balance between
polymerization and ablation can be controlled~\cite{kay,yasuda}.
It has also been shown that in plasmas the addition of
hydrogen or hydrocarbon gases  to the $CF_4$ decreases the
$F/CF_2$ ratio, thus selectively etching $Si O_2$, while 
the addition of $O_2$ or $CO_2$ increases the $F/CF_2$ ratio,
thereby selectively etching $Si$~\cite{kushner,mogab}.
In general, the addition of oxygenated species shifts the
chemistry of $CF_4$ plasmas toward etching while the addition
of hydrogenated species shifts the chemistry toward polymerization.


 It is also worthwhile to note that
in aging studies for the ATLAS muon spectrometer
the cathode cleanliness of
aluminum proportional chambers was identified as an important factor
in influencing the chamber lifetime 
even in the hydrocarbon free mixture
$Ar/CO_2/N_2$ (90:5:5)+1200~$ppm$ $H_2O$~\cite{dissert}. 
 Hydrocarbon greases from the production process are not
always completely removed 
during the cleaning procedure of the aluminum tubes.
In the case of 
incompletely cleaned tubes, sputtering by
UV-photons from wire avalanches can lead to the 
removal of non-volatile hydrocarbons from the cathode, which
then drift and stick to the wires, or can initiate the polymerization
processes in the chamber.
 Radiation tests with $Ar/CF_4$, which is expected 
to have a strong etching capability, can also confirm that 
aging effects
in an $Ar/CF_4/CH_4$ mixture are not initiated by 
trace contaminants in the gas and 
appear only when $CH_4$ is added to $Ar/CF_4$.
 


 Aging studies with an $Ar/CF_4$ (70:30) mixture were performed 
with two tube chamber cells N15,N16, 
connected in a serial gas flow.
 A pre-mixed gas was transported to the chamber by
stainless steel lines that
excluded the presence of $H_2O$ in the gas.
 The gas flow rate was set to 1.5~l/hour.
 In order to investigate the influence of photons on aging performance,
the operating voltage was set to 2.6~kV, where the rate
of afterpulses is very large~(see Fig.~\ref{fig3}).
The average current density was $\sim$~150~nA/cm. 
 During 10 days of exposure, which resulted
in a total collected charge of 70~$\frac{mC}{cm \cdot wire}$,
we have not observed either dark current
or changes in operating current.
 Further electron beam tests 
revealed no change in efficiency for both irradiated wires
(see Fig.~\ref{graf3}).
 The appearance of some wire specimens after exposure
in $Ar/CF_4$ was slightly more black than that of new ones, however, 
SEM imaging revealed clean $Au$ anode wire
surfaces, with negligible amount of $C$ found at 5~kV
SEM voltage.
The EDX spectrum at 20~kV shows only gold (see Fig.~\ref{graf7}).
 EDX analysis of irradiated cathode surfaces revealed some 
traces of fluorocarbon deposits (see Fig.~\ref{graf5}). 

{\bf{ From our results so far, no loss in performance 
of aluminum proportional chambers operated with
$Ar/CF_4$ (70:30) without water
has been observed up to collected charge 
 $\sim$~70~$\frac{mC}{cm \cdot wire}$.}}

\section{ Aging in a $CF_4/CH_4$(90:10) gas mixture}

In wire chamber operation, many studies have demonstrated 
excellent aging properties, up to 10~$\frac{C}{cm \cdot wire}$,
of $CF_4/i C_4 H_{10}$ (80:20) avalanches, which also has an ability
to etch silicon-based and hydrocarbon deposits from previously
aged gold-plated wires~\cite{opensh}. 
Aging studies have also been performed
with a $CF_4/CH_4$ (90:10) mixture and no deterioration was found
up to 1.9~$\frac{C}{cm \cdot wire}$~\cite{d0}. At the same time, 
heavy carbonaceous deposits (without incorporation of $CF_x$
fragments into the polymer) 
were formed on the gold-plated wires irradiated 
in the mixtures $CF_4/i C_4 H_{10}$ (95:5),  $CF_4/i C_4 H_{10}$
(20:80) and $CF_4/C_2 H_4$ (95:5)~\cite{chemmod}. 
Clear rapid aging effects
were also seen in honeycomb drift chambers operated
with $Ar/CF_4/CH_4$ (74:20:6) and $CF_4/CH_4$ (80:20) mixtures
in high-rate hadronic environments~\cite{stegmann,kolanoski,dissert1}.
 From our results so far, aging effects were observed
in aluminum proportional chambers irradiated with $Ar/CF_4/CH_4$
(74:20:6) and $Ar/CF_4/CH_4$ (67:30:3) mixtures (even with
the incorporation of $CF_x$ fragments into the polymer structure).
 In order to find a link between the divergent results 
obtained in $CF_4$/hydrocarbon and $Ar/CF_4/CH_4$ mixtures,
we performed radiation tests with $CF_4/CH_4$ (90:10) 
and $CF_4/CH_4$ (80:20) gases in the high-rate HERA-B environment. 



 Fig.~\ref{graf6}a shows tube chamber efficiency 
for $CF_4/CH_4$ (90:10) and $CF_4/CH_4$ (80:20) gases,
measured  in an electron beam before the aging run, 
as a function of high voltage.
 Aging studies with $CF_4/CH_4$ (90:10) + 600 $ppm$ $H_2O$ mixture
have been carried out on a set of
two chamber cells N1,N2 connected in a serial gas flow.
 We have employed the flow rate of 1.5~l/hour.
 Wires were operated at a high voltage of
3.0~kV, which led to an average current density of 200~$nA/cm$.
 The total accumulated charge at the end of the run was 
$\sim$~370~$\frac{mC}{cm \cdot wire}$.
After the aging test, scanning along the wires in an
electron beam showed no loss in performance
for most wire specimens, except a region
($\sim$~4~cm) of wire N2 near one of the chamber endcaps, where a
large efficiency drop was found~(see Fig.~\ref{graf4}).
However, the reason for this inefficiency remained unclear due to 
technical reasons.  
  Fig.~\ref{graf6}b shows SEM imaging of wire N2's central region, 
placed downstream of the damaged area in the 
direction of gas flow,
where the EDX analysis (at 20~kV SEM voltage)
revealed only $Au$ signal. In the 
EDX spectrum of the irradiated cathode we did not observe
any significant layer of deposits on the surface (see Fig.~\ref{graf5}).

\section{ Aging in a $CF_4/CH_4$(80:20) gas mixture}

We have studied the aging properties of a 
$CF_4/CH_4$(80:20) mixture in two ways: 
1) with 600~$ppm$ $H_2O$ and 2) without additives.

 Initially, radiation tests with a
$CF_4/CH_4$(80:20) + 600~$ppm$ $H_2O$
mixture were carried out on a set of 
three chamber cells - N7,N8,N9, connected in a serial gas flow.
We have employed a flow rate of 1.5~l/h.
 Wires were operated at a high voltage of 3.0~kV, 
which led to an average current density $\sim$~150~nA/cm. 
 During one month of exposure, which resulted 
in a collected charge of $\sim$~170~$\frac{mC}{cm \cdot wire}$,
we have not observed either dark current or
changes in operating current.
 However, scanning along the wires in an electron beam
revealed a loss in efficiency for all irradiated wires.
 Fig.~\ref{graf22} shows efficiency profiles for wires N7,N8,N9
after an accumulated charge   $\sim$~170~$\frac{mC}{cm \cdot wire}$.
It is of  special interest that aging effects
in a $CF_4/CH_4$(80:20) + 600~$ppm$ $H_2O$ 
mixture only slightly increases
in the direction of the serial gas flow (from wire N7 to N9) and
are much more severe in the center of the wires (x~$\sim$~250~cm), 
where the radiation intensity was the highest (see Fig.~\ref{graf22}).



 At the conclusion of the tests anode wires and cathodes were
disassembled from the chamber and inspected with SEM and EDX.
Fig.~\ref{pic7} shows micrographs of wire N8 at 20~kV and 5~kV
SEM voltages. 
 While in the EDX spectrum at 20~kV there is no 
significant concentration of elements other than the wire material,
 the EDX spectrum at 5~kV is dominated by an intense $C$ 
peak (see Fig.~\ref{pic7}).
However, these deposits are far less substantial than
those observed in the $Ar/CF_4/CH_4$ mixture.
 Only negligible amounts of $C$, $O$ and $F$ were  identified
on irradiated cathodes after exposure in a 
$CF_4/CH_4$(80:20) + 600~$ppm$ $H_2O$ mixture (see Fig.~\ref{graf5}).
 
In the second run, 
all polyamid tubes were exchanged for stainless steel ones, 
and aging studies with a $CF_4/CH_4$(80:20) mixture without water
were performed on another set of the wires - N11,N12,N13.
 All operating conditions were kept as in the previous
run (gas flow - 1.5~l/hour, high voltage - 3.0~kV).
Due to the smaller radiation intensity for wires N11,N12,N13,
the average current density was lower $\sim$~100~nA/cm.
 During exposure, we 
have not noticed any operational degradation
of the wire performance.
After 10 days of irradiation,
which resulted in a total collected charge of 
$\sim$~40~$\frac{mC}{cm \cdot wire}$, 
studies in an electron beam have shown that all three wires
are fully efficient.

{\bf{ In contrast to the $Ar/CF_4/CH_4$ mixture,
aging tests of aluminum proportional chambers filled with
$CF_4/CH_4$ (90:10) and $CF_4/CH_4$ (80:20) did not show either
a rapid gain reduction at the level of accumulated charges 
$\sim$~30-100~$\frac{mC}{cm \cdot wire}$, or the
appearance of the dark current.
No dependence of lifetime on water addition and size of
irradiation area was observed for $CF_4/CH_4$ gases. 

 Using $CF_4/CH_4$ (90:10)+ 600 $ppm$ $H_2O$, a stable lifetime
of up to 370~$\frac{mC}{cm \cdot wire}$ was achieved for two
wires, except for one wire 
specimen ($\sim$~4~cm out of 50~cm), 
where a significant gain reduction was detected. The origin
of this aging is not understood.

 Using $CF_4/CH_4$ (80:20)+ 600 $ppm$ $H_2O$, a gain reduction
was observed after a radiation dose of
170~$\frac{mC}{cm \cdot wire}$ for all three irradiated wires.
 SEM analysis confirmed the  presence of a thin layer of a carbonaceous
deposit on the anode wires. At the same time, 
no loss in performance was observed
for three wires operated with a $CF_4/CH_4$ (80:20) mixture
without water up to 40~$\frac{mC}{cm \cdot wire}$.
}}

\section{ Aging in an $Ar/CF_4/CH_4$ (74:20:6) gas mixture}

 From our results reported in~\cite{max},
rapid aging effects were observed during
irradiation of the tube chamber filled with an
$Ar/CF_4/CH_4$ (74:20:6) mixture in a 100~MeV $\alpha$-beam
of the cyclotron in the research center "Forschungszentrum
Karlsruhe GmbH". Therefore, this mixture was clearly ruled out for 
the use in the muon detector on the basis of it's aging properties.
However, radiation tests with 
$Ar/CF_4/CH_4$ (74:20:6) were also performed in the HERA-B environment.
As operating parameters were not kept constant during the 
aging run, these results can be only used  
for a qualitative picture of the aging behavior of 
$Ar/CF_4/CH_4$ (74:20:6) mixture under high-rate irradiation.


The experimental setup and operating conditions during the 
aging studies were the following.
Tube chamber with 16 drift cells and a cross cell
exactly as for the production version, but of a shorter length of 50~cm,
was placed between the electromagnetic calorimeter and muon absorber,
as shown in Fig.~\ref{tabl1}.
 A pre-mixed gas was transported 
by a 150~m stainless steel tube followed by a polyamid tube 
connected directly to the chamber inlet. 
However, the chamber outlet was not connected to a gas chromatograph, 
which left us unable to measure the level of species
($N_2$, $O_2$ $H_2 O$) in the effluent gas stream.
 During the aging run,  
the chamber was operated at several high voltages 
in the range 2.3~-~2.5~kV~(see Fig.~\ref{fig3}), 
which led to average currents of
80-300~$\mu A$, measured from 16 anode wires. The gas flow
also varied during the tests from 3~l/h up to 15~l/h.
 By monitoring the chamber current 
the average collected charge was determined.  
Similar to the performance of aged wires in an $Ar/CF_4/CH_4$ (67:30:3) 
mixture,
the 'switch-on' current behavior and the dependence of the current on 
the gas flow rate also appeared
during operation with $Ar/CF_4/CH_4$ (74:20:6) in HERA-B,
indicating the onset of aging effects.

It should also be noted that in order to study
the dependence of the anode current on the gas flow, 
chamber flushing was stopped a few times 
for intervals of approximately 20 minutes.
 Irradiation of the chamber was halted when the average collected charge
for the wires had reached approximately 200~$\frac{mC}{cm \cdot wire}$.
 When, at the conclusion of the tests, the
 chamber was opened for inspection,
surface deposits were found on the anode wires.
 Fig.~\ref{gr51} shows typical micrographs of deposits for
two wire specimens. EDX analysis of these 
wires revealed the presence of a polymer coating, consisting of
$C$, $F$ (and most probably $O$) elements.
  
{\bf{ It is of special interest that, as in our previous studies
using a 100~MeV $\alpha$-beam with an $Ar/CF_4/CH_4$ (74:20:6) mixture
and results with $Ar/CF_4/CH_4$ (67:30:3) in
the HERA-B environment,
these tests also revealed the presence of fluorine in 
the anode wire deposits.}}

\section{ Summary }

High energy and luminosity experiments 
pose a new challenge for the construction and operation
of large area gaseous detectors.
The HERA-B muon system, with a total gas volume of 8~$m^3$,
puts rather stringent constraints
on the gaseous medium to be used: extremely low aging, 
high resistance to sparking, good transport properties (high drift 
velocity $v_d$, convenient operating electric field E/p),
good chemical properties (non-flammable, non-poisonous).
 Under these constraints only a limited choice 
of gases can be used and, moreover, the new generation of
experiments demand a higher radiation hardness 
than available from conventional mixtures
($ \leq 1~\frac{C}{cm \cdot wire}$). 

About twenty years ago, $CF_4$ was proposed as a 
most attractive candidate for the
wire chamber operation in high rate environments.
 This is primarily due to the high drift velocity, high primary 
ionization density and low electron diffusion of 
$CF_4$~\cite{fast1,fischer,schmidt1,cf41}. Many
studies also have demonstrated an excellent aging performance of
several $CF_4$-based mixtures, which suppress aging effects 
up to an exposure of 
$\sim~10~\frac{C}{cm \cdot wire}$~\cite{kadyk1,opensh1}.  
Within the broad spectrum of gases, there are no mixtures
without $CF_4$ that are able to tolerate such radiation doses.
From the viewpoint of plasma polymerization
the $CF_4$ molecule is an ideal source for a variety of reactive
neutral and ionic fragment atoms and molecules formed in either 
the ground or excited states which are largely responsible
for surface reactions in various etching and deposition applications.
It is believed that, when $CF_4$ dissociates in the gaseous discharges
into highly reactive 
$CF_x$ and $F$ radicals, especially the elemental fluorine
is very effective in suppressing polymerization in the wire chamber.
Actually, in a plasma environment $CF_4$-based gases are used for 
both etching and deposition processes, 
the distinction being made by the gas and its concentration with which 
$CF_4$ is mixed~\cite{yasuda}. 
 The high radiation load of the HERA-B experiment
and the requirement of a  
fast signal collection within the 96~ns time interval
between two consecutive bunch crossings, led us to
consider one of the  $Ar/CF_4$-based mixtures:
$Ar/CF_4/CH_4$ (74:20:6), $Ar/CF_4/CH_4$ (67:30:3) 
$Ar/CF_4/CH_4$ (65:30:5) for the muon detector operation.

 The aging performance of aluminum proportional chambers 
has been investigated for five
different $CF_4$-containing mixtures in three different 
experimental setups: laboratory conditions ($Ru^{106}$ and $Fe^{55}$
sources),
a 100~MeV $\alpha$-beam and the high-rate HERA-B environment.
 The size of the irradiation zone varied in the tests 
from 1~cm  up to 500~cm. Some of the results obtained 
in the framework of these studies have already been 
published~\cite{max}.
 In the following the summary of aging studies as well as possible
application of these results to the construction of a large area 
gaseous detectors for operation 
in high rate environments is presented.
 The effect of added $CF_4$ on the aging performance
of wire chambers and some possible chemical mechanisms
in the $CF_4$-containing
gaseous discharges will be discussed in a companion paper~\cite{max2}.

1. The aging rate of aluminum proportional chambers
filled with $Ar/CF_4/CH_4$ (74:20:6) and 
$Ar/CF_4/CH_4$ (67:30:3) mixtures
differs by more than two orders of magnitude for laboratory tests
with radioactive sources and in the high-rate radiation
environments. 
These results clearly indicate that
the aging rate depends on a particular set of operating
parameters and, therefore, 
from the accumulated charge alone it is not possible
to combine the data from the different experimental setups into one 
consistent model.
 In a view of the aging results presented here it is important to 
note that the initial stage of radiation tests 
usually performed in the laboratory may
not offer full information, needed to give an estimation about 
the lifetime of the real detector. 

 2. Aging effects in  tube proportional chambers 
filled with $Ar/CF_4/CH_4$ (74:20:6), 
$Ar/CF_4/CH_4$ (67:30:3) mixtures were observed after irradiation
in a 100~MeV $\alpha$-beam and in the high-rate HERA-B environment.
 For operation in HERA-B the aging behavior of the
$Ar/CF_4/CH_4$ (67:30:3) mixture was found to depend
on the high voltage and the size of the irradiation area.
 SEM analysis of the aged anode wires, operated with a 
gas gain $>10^4$, revealed a polymer matrix, consisting of 
$C$ and $F$ elements ($H$ is not detectable).
Anode aging
was also accompanied by the appearance of a thin layer of  
polymers on the cathode, consisting of $C$, $F$ and $O$.
These surface deposits on the electrodes
resulted in a gain reduction and in the appearance of a dark current
during operation in HERA-B.
In contrast to deposits formed 
at a gas gain $>10^4$, one of the anode wires 
operated at a gas amplification factor
$<10^4$ had only trace carbonaceous deposits, without 
incorporation of $F$ into the polymer structure.
 Although, 
addition of $H_2O$ results in the suppression of the polymerization 
effects in the $Ar/CF_4/CH_4$ (67:30:3) mixture,
 our results clearly indicate a problem for the use of $Ar/CF_4/CH_4$
mixture in the muon detector at the HERA-B experiment.

In the wire chamber operation
fluorocarbon deposits were usually not observed on the 
$Au/W$ anode wires~\cite{chemmod}.
 This was mainly attributed to the fact that 
$CF_4$ can not polymerize without breaking of the $C$-$F$
bond. Once fluorine is split off, however, it's ablative effect 
becomes predominant. 
 From the results of tube chamber operation with
$Ar/CF_4/CH_4$ (74:20:6) and  $Ar/CF_4/CH_4$ (67:30:3) mixtures, 
it is evident that  
fluorine may be also incorporated into a polymer matrix
on the anode wires.
 At the same time, different aging performance of 
$CF_4/CH_4$ (90:10) and $CF_4/CH_4$ (80:20) gases, where
fluorine was not detected on the wires lead us to the conclusion,
that the aging properties can not be solely explained on the basis of
the molecule ratios of the gases, without taking into
account the actual discharge conditions.
The change in operating parameters 
(gas gain, area of irradiation, gas flow)
could be one of the possible explanations of the difference in aging 
behavior between $Ar/CF_4/CH_4$ (67:30:3) and $CF_4/CH_4$ (90:10),
which have nearly identical ratio of $CF_4/CH_4$  molecules
in the mixture.

  Strong dependence of the aging rate on 
high voltage and the size of the irradiation area
has been also reported in~\cite{dissert,kolly1}.
 Since presumably the high voltage is not the physical quantity
which is directly responsible for aging in wire chambers, 
the avalanche-related aging effects can be classified as  
depending on both the total charge of the avalanche (avalanche size)
and on the gas amplification, which is related 
to the mean electron energy in the avalanche and, therefore,
to different excitation and ionization phenomena.
 Thus, the increase in high voltage could either result  
in a significantly larger pulse charges, especially if 
 self-quenching streamers appears,
or in discharges (glow discharges or sparks). The increase
in high voltage could also 
initiate the production of new reactive species, 
or produce them at a much larger rates thus promoting
the polymer formation.

The main principles of traditional plasma chemistry 
(low pressure, rf) are generally used to predict chemical reactions 
in the wire avalanches~\cite{chemmod,vavra,kadyk}.
The dependence of the polymer formation on the 
energy input level is well established in plasma polymerization.
 The 'fragmentation' of molecules in a plasma is highly dependent 
on the structure of the starting material and the 
discharge conditions. Nearly all organic compounds
regardless of their chemical nature can be polymerized. Particularly,
polymers consisting of $C$ and $F$ elements 
were produced by the polymerization of the etching gas 
$CF_4$~\cite{arikado, martz, gasflow1}.
 In general, considerable fragmentation of the initial gas or
rearrangement of atoms occurs in plasma and 
the extent of the process and the 
dominating mechanisms vary with the type of gas and the discharge
conditions.
Because of this aspect, the structure
of plasma polymers formed from the same 
monomer is highly dependent on the actual conditions of plasma 
polymerization, in particular, the energy input level,
the size (cross-sectional area) of a tube or reactor and 
even on the position within the reactor~\cite{yasuda}.

The experimental data presented here indicate that
the balance between polymer formation and 
ablation in the wire avalanches
could depend not only on the chemical nature of the gas 
mixture, but also on the discharge conditions, 
particularly the energy density. The trends found in these studies
require to simulate the final radiation conditions of the experiment
as well as possible in order to choose the mixture
for operation in the high-rate environments.

3. Recent studies have shown also the dependence of aging 
properties on the size of the irradiation area, in particular, an
increase of the aging rate in the direction of the serial gas 
flow~\cite{dissert}.
Another example is that
developments of anode aging and Malter effects in 
honeycomb drift chambers were not so pronounced in short
$\sim$~10~$cm$ chambers as in  long 
$\sim$~100~$cm$ chambers~\cite{dissert1}.
  These observations seem to be the most critical when
trying to extrapolate the aging behavior from small 
to large areas of irradiation. 
In a naive approach these effects can be explained as 
follows. Upon the repeated gaseous discharges,
polymer-forming species are produced and 
diffuse in the direction of the gas flow. 
During this diffusion, some of the long-lived radicals may 
migrate and react with active polymer fragments even outside 
the avalanche region, thus resulting in a 
further growing of the polymerized chains.

In plasma polymerization of $CF_4$ the degree of conversion
of the starting material was found to be 
a monotonic function of the residence time in the discharge 
zone~\cite{gasflow}. 
However, the extent to which this result can be applied
to the wire chamber operation, where the 
production of reactive species is mostly confined
to a small region ($\sim 50~\mu m$) 
near the anode wire, is not clearly understood.
 It is important to note that due to the increased
aging effects in the direction of the gas flow 
it is worthwhile to avoid gas systems, that supplies
many chambers in a serial flow.

4.  The hydrocarbon free gases are much more resistant 
to aging than gases with hydrocarbons, especially
for operation in a high-rate radiation environment.
No loss in performance of the aluminum 
chambers was observed in the hydrocarbon free gases 
after an accumulated charge of 700~$\frac{mC}{cm \cdot wire}$ in 
an $Ar/CF_4/CO_2$ (65:30:5) + 1000~$ppm$ $H_2O$ and of 
70~$\frac{mC}{cm \cdot wire}$ in an $Ar/CF_4$ (70:30) mixtures
in the real HERA-B environment.

5. In order to successfully operate gaseous detectors
at severe radiation conditions,
one should certify the gas purity and chemical reactivity
of the various materials in order to avoid the presence of 
'bad' molecules in contact with the gas system. 
 Furthermore, the effects of the cathode materials on the aging 
performance should be investigated, 
especially for $CF_4$-containing  mixtures, since
polymers were found on the non-gold plated tungsten wires,
after exposure in a strongly etching mixture
$CF_4/iC_4 H_{10}$ (80:20)~\cite{chemmod}.
A detailed summary of the influence of
commonly used materials on aging properties 
may be found in ~\cite{kadyk,romaniouk}. 

6. Since the aging effects are of a statistical nature,
it is of primary importance 
that the radiation tests should be carried out on
a set of wires irradiated under identical conditions. 
Such a study will allow for the exclusion of statistical fluctuations
in the aging performance and provide a reliable estimation
about the detector lifetime.
 
Finally, many chemical processes  are expected to occur 
simultaneously in the gaseous discharges 
surrounding the wire and therefore
a quantitative description of aging effects, which 
would require as a minimum, a detailed analysis of all gas phase 
and gas surface reaction products, is not possible. 
In addition, many of the reactions may be extremely 
sensitive to the nature and purity of the gas mixture, different additives
and trace contaminants, construction materials, the actual
geometry of the electrodes, the gas flow, the irradiation intensity,
the size of the irradiation area, the particle species...
 Since, the present state of knowledge does not allow one to formulate
a complete set of recommendations of how to prevent the aging effects
in wire chambers,
it is important to study the aging performance
of the mixture in the conditions as closely as possible to real ones.
 In a view of the aging results discussed here, one can also see
that in wire chamber operation,
the presence of large amounts of $CF_4$ in the mixture does not 
necessarily ensure good aging performance.



\section*{Acknowledgments.}

 We would like to thank Dr. G. Bohm and Wildau
Politechnik Institute for the possibility to analyse the wires.
 We thank to K. Reeves and S. Aplin and W. Hulsbergen
 for reading and correcting this manuscript.

 This work was partly supported by the 
Deutsches Elektronen-Synchrotron (DESY), 
by the Alexander von Humboldt-Stiftung and Max Planck Research Award.

\newpage

\begin{figure}
\caption{ Efficiency profiles along wires N15,N16 
after an accumulated charge $\sim$~70~$\frac{mC}{cm \cdot wire}$
in $Ar/CF_4$ (70:30) mixture. For comparison,
efficiency for reference wire N14 is also shown.
(Measurements were performed at high voltage 2.4~kV with
$Ar/CF_4/CH_4$ (67:30:3) mixture).
    \label{graf3}}
\end{figure}

\begin{figure}
\caption{ SEM micrographs of wire N16, irradiated in
an $Ar/CF_4$(70:30) mixture up to an accumulated 
charge $\sim$~70~$\frac{mC}{cm \cdot wire}$, show the different
'EDX' sensitivity at 20~kV and 5~kV.
  \label{graf7}}
\end{figure}

\begin{figure}
\caption{ EDX spectroscopy of $Al$ cathodes, irradiated in the
mixtures: $Ar/CF_4$ (70:30),
$CF_4/CH_4$(90:10) + 600~$ppm$ $H_2O$,
$CF_4/CH_4$(80:20) + 600~$ppm$ $H_2O$.
  \label{graf5}}
\end{figure}

\begin{figure}
\caption{ Efficiency profiles along wires N1,N2 measured
in an electron beam,
after an accumulated charge of $\sim$~370~$\frac{mC}{cm \cdot wire}$ 
in an $CF_4/CH_4$ (90:10) + 600 $ppm$ $H_2O$ mixture. These results are to
be compared with efficiency of reference wire (N3) $\ge$~99~$\%$ 
at 2.9~kV.
\label{graf4}}
\end{figure}

\begin{figure}
\caption{ a) Tube chamber efficiency as a function of high voltage;
   b) SEM micrograph of wire N2 central region.
   This region was subjected to a radiation dose of 
   $\sim$~370~$\frac{mC}{cm \cdot wire}$ 
   in an $CF_4/CH_4$ (90:10) + 600 $ppm$ $H_2O$ mixture.
   EDX analysis revealed only $Au$ signal.
\label{graf6} }
\end{figure}

\begin{figure}
\caption{ Efficiency profile along wires (N7,N8,N9), measured in an
 electron beam, 
 after an accumulated charge $\sim$~170~$\frac{mC}{cm \cdot wire}$ 
  in an $CF_4/CH_4$(80:20) + 600~$ppm$ $H_2O$. These results are to be 
  compared with efficiency of reference wire (N10) $\ge$~99~$\%$ at 2.9~kV.
 \label{graf22}}
 \end{figure}

\begin{figure}
\caption{ SEM micrographs of wire N8, irradiated in an
$CF_4/CH_4$(80:20) + 600~$ppm$ $H_2O$ mixture up to an accumulated 
charge  of $\sim$~170~$\frac{mC}{cm \cdot wire}$, show the different
EDX sensitivity at 20~kV and 5~kV.
\label{pic7}}
\end{figure}

\begin{figure}
\caption{ SEM micrographs of two specimens, taken 
from different chamber wires, after irradiation in an 
$Ar/CF_4/CH_4$ (74:20:6) mixture. In the EDX spectrum, lines
corresponding to $C$, $F$ (and most probably $O$) are detected.
  \label{gr51}}
\end{figure}

\begin{figure}
\caption{ a) Tube chamber efficiency 
as a function of high voltage; b) Singles counting rate as a 
function of high voltage.
  \label{fig3}}
\end{figure}

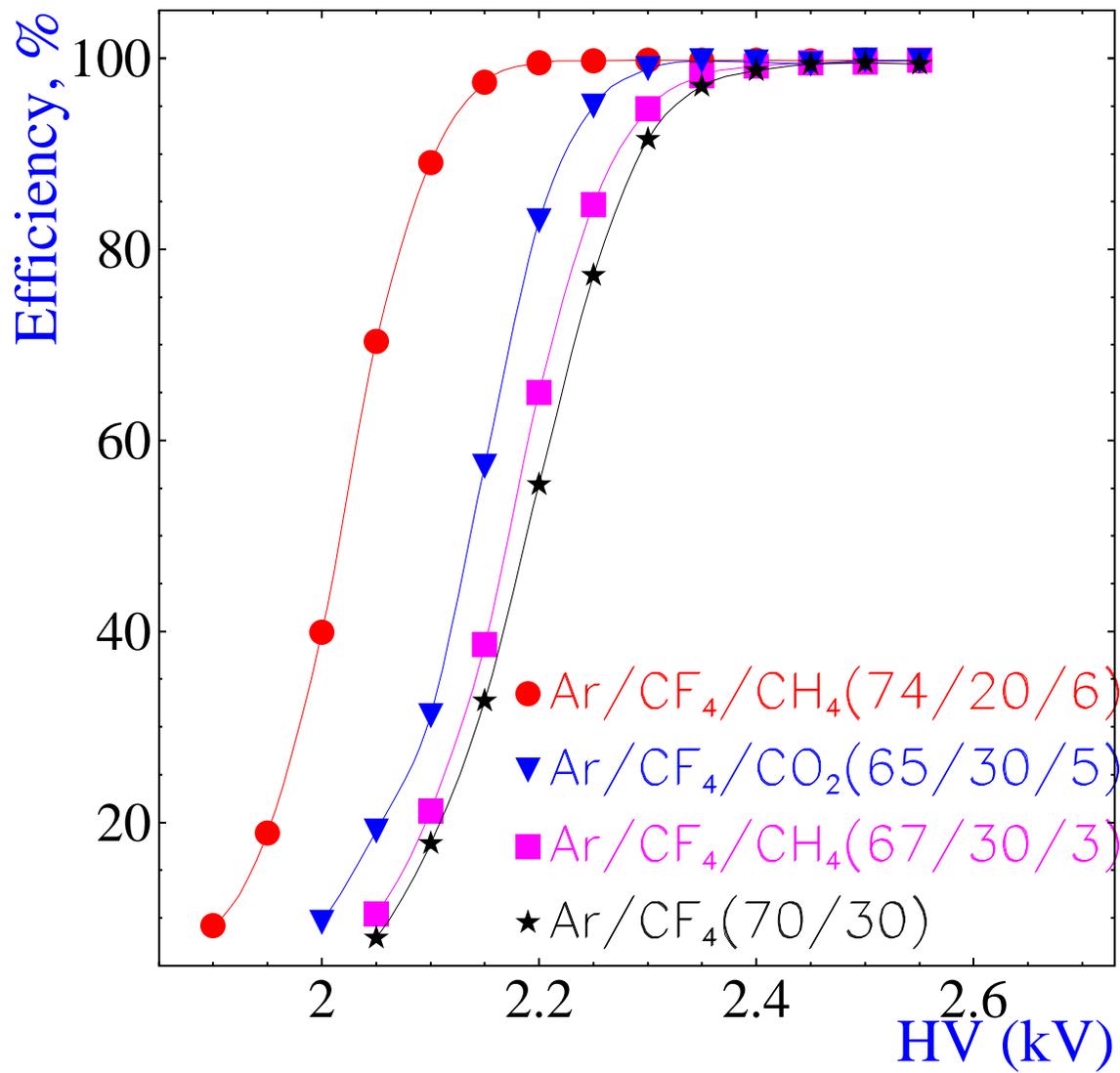
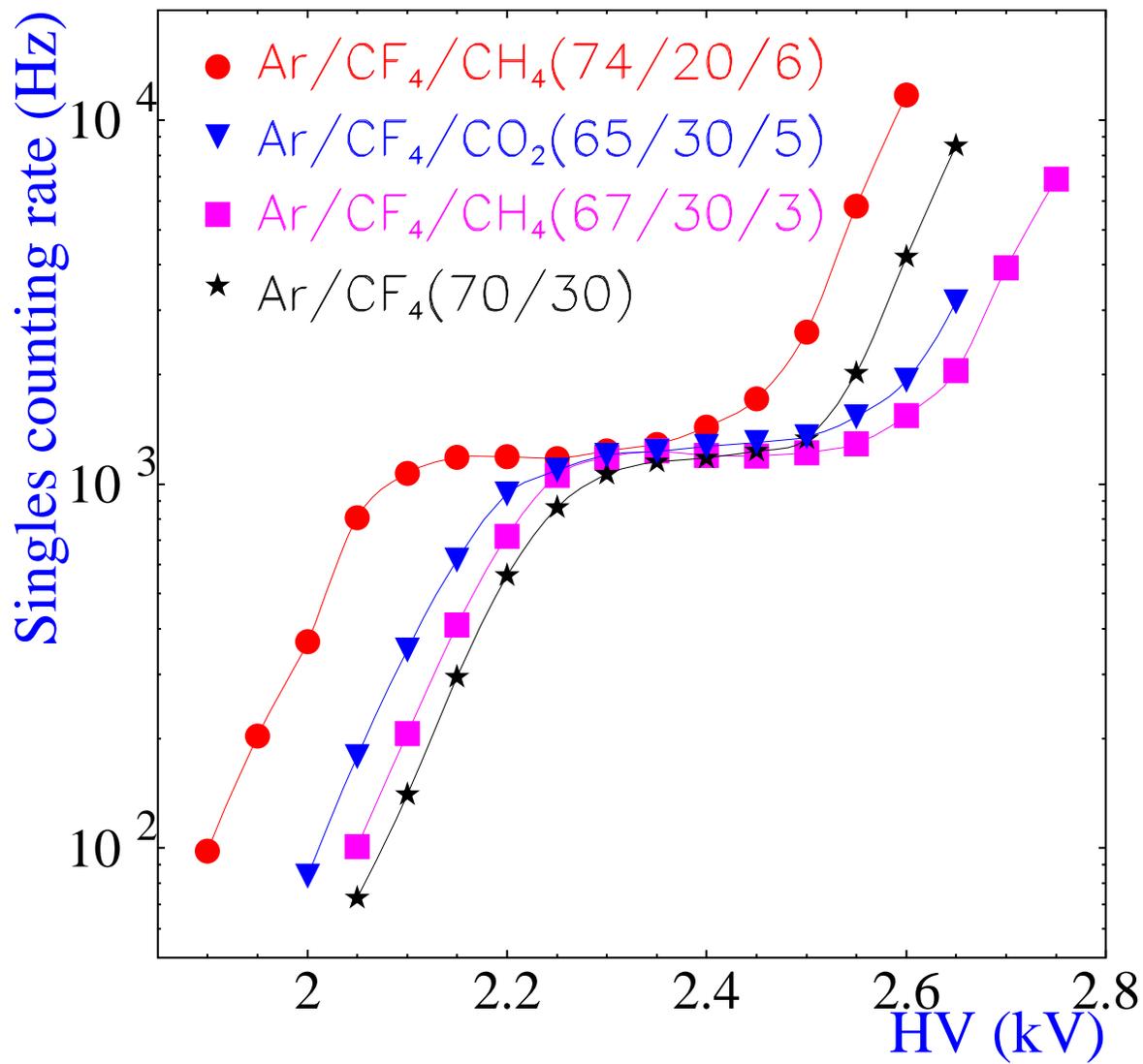
\end{document}